\documentstyle[twoside,epsfig]{article}
\catcode`\@=11
\long\def\@makefntext#1{
\protect\noindent \hbox to 3.2pt {\hskip-.9pt  
$^{{\eightrm\@thefnmark}}$\hfil}#1\hfill}		

\def\@makefnmark{\hbox to 0pt{$^{\@thefnmark}$\hss}}	
	
\def\ps@myheadings{\let\@mkboth\@gobbletwo
\def\@oddhead{\hbox{}
\rightmark\hfil\eightrm\thepage}   
\def\@oddfoot{}\def\@evenhead{\eightrm\thepage\hfil
\leftmark\hbox{}}\def\@evenfoot{}
\def\sectionmark##1{}\def\subsectionmark##1{}}

\oddsidemargin=\evensidemargin
\newcounter{sectionc}\newcounter{subsectionc}\newcounter{subsubsectionc}
\renewcommand{\section}[1] {\vspace{12pt}\addtocounter{sectionc}{1} 
\setcounter{subsectionc}{0}\setcounter{subsubsectionc}{0}\noindent 
	{\tenbf\thesectionc. #1}\par\vspace{5pt}}
\renewcommand{\subsection}[1] {\vspace{12pt}\addtocounter{subsectionc}{1} 
	\setcounter{subsubsectionc}{0}\noindent 
	{\bf\thesectionc.\thesubsectionc. {\kern1pt \bfit #1}}\par\vspace{5pt}}
\renewcommand{\subsubsection}[1] {\vspace{12pt}\addtocounter{subsubsectionc}{1}
	\noindent{\tenrm\thesectionc.\thesubsectionc.\thesubsubsectionc.
	{\kern1pt \tenit #1}}\par\vspace{5pt}}
\newcommand{\nonumsection}[1] {\vspace{12pt}\noindent{\tenbf #1}
	\par\vspace{5pt}}

\newcounter{appendixc}
\newcounter{subappendixc}[appendixc]
\newcounter{subsubappendixc}[subappendixc]
\renewcommand{\thesubappendixc}{\Alph{appendixc}.\arabic{subappendixc}}
\renewcommand{\thesubsubappendixc}
	{\Alph{appendixc}.\arabic{subappendixc}.\arabic{subsubappendixc}}

\renewcommand{\appendix}[1] {\vspace{12pt}
        \refstepcounter{appendixc}
        \setcounter{figure}{0}
        \setcounter{table}{0}
        \setcounter{lemma}{0}
        \setcounter{theorem}{0}
        \setcounter{corollary}{0}
        \setcounter{definition}{0}
        \setcounter{equation}{0}
        \renewcommand{\thefigure}{\Alph{appendixc}.\arabic{figure}}
        \renewcommand{\thetable}{\Alph{appendixc}.\arabic{table}}
        \renewcommand{\theappendixc}{\Alph{appendixc}}
        \renewcommand{\thelemma}{\Alph{appendixc}.\arabic{lemma}}
        \renewcommand{\thetheorem}{\Alph{appendixc}.\arabic{theorem}}
        \renewcommand{\thedefinition}{\Alph{appendixc}.\arabic{definition}}
        \renewcommand{\thecorollary}{\Alph{appendixc}.\arabic{corollary}}
        \renewcommand{\theequation}{\Alph{appendixc}.\arabic{equation}}
        \noindent{\tenbf Appendix \theappendixc #1}\par\vspace{5pt}}
\newcommand{\subappendix}[1] {\vspace{12pt}
        \refstepcounter{subappendixc}
        \noindent{\bf Appendix \thesubappendixc. {\kern1pt \bfit #1}}
	\par\vspace{5pt}}
\newcommand{\subsubappendix}[1] {\vspace{12pt}
        \refstepcounter{subsubappendixc}
        \noindent{\rm Appendix \thesubsubappendixc. {\kern1pt \tenit #1}}
	\par\vspace{5pt}}

\newcommand{\textlineskip}{\baselineskip=13pt}
\newcommand{\smalllineskip}{\baselineskip=10pt}

\def\eightcirc{
\begin{picture}(0,0)
\put(4.4,1.8){\circle{6.5}}
\end{picture}}
\def\eightcopyright{\eightcirc\kern2.7pt\hbox{\eightrm c}} 

\newcommand{\copyrightheading}[1]
	{\vspace*{-2.5cm}\smalllineskip{\flushleft
	{\footnotesize International Journal of Modern Physics A, #1}\\
	{\footnotesize $\eightcopyright$\, World Scientific Publishing
	 Company}\\
	 }}

\def\abstracts#1#2#3{{
	\centering{\begin{minipage}{4.5in}\baselineskip=10pt\footnotesize
	\parindent=0pt #1\par 
	\parindent=15pt #2\par
	\parindent=15pt #3
	\end{minipage}}\par}} 

\renewenvironment{thebibliography}[1]
	{\frenchspacing
	 \ninerm\baselineskip=11pt
	 \begin{list}{\arabic{enumi}.}
	{\usecounter{enumi}\setlength{\parsep}{0pt}
	 \setlength{\leftmargin 12.7pt}{\rightmargin 0pt} 
	 \setlength{\itemsep}{0pt} \settowidth
	{\labelwidth}{#1.}\sloppy}}{\end{list}}

\newcounter{itemlistc}
\newcounter{romanlistc}
\newcounter{alphlistc}
\newcounter{arabiclistc}

\newcommand{\fcaption}[1]{
        \refstepcounter{figure}
        \setbox\@tempboxa = \hbox{\footnotesize Fig.~\thefigure. #1}
        \ifdim \wd\@tempboxa > 5in
           {\begin{center}
        \parbox{5in}{\footnotesize\smalllineskip Fig.~\thefigure. #1}
            \end{center}}
        \else
             {\begin{center}
             {\footnotesize Fig.~\thefigure. #1}
              \end{center}}
        \fi}

\newcommand{\tcaption}[1]{
        \refstepcounter{table}
        \setbox\@tempboxa = \hbox{\footnotesize Table~\thetable. #1}
        \ifdim \wd\@tempboxa > 5in
           {\begin{center}
        \parbox{5in}{\footnotesize\smalllineskip Table~\thetable. #1}
            \end{center}}
        \else
             {\begin{center}
             {\footnotesize Table~\thetable. #1}
              \end{center}}
        \fi}

\def\@citex[#1]#2{\if@filesw\immediate\write\@auxout
	{\string\citation{#2}}\fi
\def\@citea{}\@cite{\@for\@citeb:=#2\do
	{\@citea\def\@citea{,}\@ifundefined
	{b@\@citeb}{{\bf ?}\@warning
	{Citation `\@citeb' on page \thepage \space undefined}}
	{\csname b@\@citeb\endcsname}}}{#1}}

\newif\if@cghi
\def\cite{\@cghitrue\@ifnextchar [{\@tempswatrue
	\@citex}{\@tempswafalse\@citex[]}}
\def\citelow{\@cghifalse\@ifnextchar [{\@tempswatrue
	\@citex}{\@tempswafalse\@citex[]}}
\def\@cite#1#2{{$\null^{#1}$\if@tempswa\typeout
	{IJCGA warning: optional citation argument 
	ignored: `#2'} \fi}}

\def\pmb#1{\setbox0=\hbox{#1}
	\kern-.025em\copy0\kern-\wd0
	\kern.05em\copy0\kern-\wd0
	\kern-.025em\raise.0433em\box0}

\def\fnt#1#2{\footnotetext{\kern-.3em
	{$^{\mbox{\scriptsize #1}}$}{#2}}}

\def\fpage#1{\begingroup
\voffset=.3in
\thispagestyle{empty}\begin{table}[b]\centerline{\footnotesize #1}
	\end{table}\endgroup}

\def\runninghead#1#2{\pagestyle{myheadings}
\markboth{{\protect\footnotesize\it{\quad #1}}\hfill}
{\hfill{\protect\footnotesize\it{#2\quad}}}}
\font\tenrm=cmr10
\font\tenit=cmti10 
\font\tenbf=cmbx10
\font\bfit=cmbxti10 at 10pt
\font\ninerm=cmr9

\font\eightrm=cmr8

\def\qed{\hbox{${\vcenter{\vbox{			
   \hrule height 0.4pt\hbox{\vrule width 0.4pt height 6pt
   \kern5pt\vrule width 0.4pt}\hrule height 0.4pt}}}$}}

\begin{document}

\runninghead{Measurement of the $\bar{B}^0$ and $B^-$ Meson Lifetimes in ALEPH}
{Measurement of the $\bar{B}^0$ and $B^-$ Meson Lifetimes in ALEPH}

\normalsize\textlineskip
\thispagestyle{empty}
\setcounter{page}{1}

\copyrightheading{}			

\vspace*{0.88truein}

\fpage{1}
\centerline{\bf MEASUREMENT OF THE ${\rm B^0}$ and ${\rm B}^-$ MESON LIFETIMES IN ALEPH}
\vspace*{0.37truein}
\centerline{\footnotesize KAY H\"UTTMANN (for the ALEPH Collaboration)}
\vspace*{0.015truein}
\centerline{\footnotesize\it Max-Planck-Institut f\"ur Physik, F\"ohringer~Ring~6, 
D-80805 Munich, Germany}

\vspace*{0.21truein}
\abstracts{
The lifetimes of the ${\rm \bar{B}^0}$ and ${\rm B}^-$ meson lifetimes are
measured using data recorded on the Z peak with the ALEPH detector at LEP.
An improved analysis based on
partially reconstructed ${\rm \bar{B}}^0 \to {\rm D}^{*+} \ell^- \bar{\nu}$ and 
${\rm B}^- \to {\rm D}^0 \ell^- \bar{\nu}$ decays is presented.}{}{}


\vspace*{1pt}\textlineskip	
\section{Introduction}	
\vspace*{-0.5pt}
\noindent
Measurements of the individual $b$ hadron lifetimes provide an important
test of our understanding of $b$ hadron decay dynamics beyond the simple 
spectator model.
Using the heavy quark expansion formalism, the lifetime hierarchy
of $b$ hadrons is predicted to be $\tau_{\Lambda_b} < \tau_{\rm \bar{B}^0}
\sim \tau_{{\rm B}^0_s} < \tau_{\rm B^-}$, with differences at 
the level of a few percent. Precise measurements are therefore needed in
order to test the theory.

This paper reports an improved measurement of the ${\rm \bar{B}^0}$ and 
${\rm B}^-$ lifetimes with the ALEPH detector at LEP based on ${\rm D^0}$-
and ${\rm D}^{*+}$-lepton samples from semileptonic B decays \cite{Paper}.  
The LEP1 data sample was recently reprocessed using refined reconstruction 
algorithms. The main improvements concern the track reconstruction and 
particle identification, resulting in enhanced reconstruction efficiencies 
for charmed mesons ranging from 10 to $30\%$ with respect to a previous 
analysis \cite{OldPaper}.

\section{Event Selection and B Reconstruction}
Semileptonic ${\rm \bar{B}^0}$ and ${\rm B}^-$ decays are partially
reconstructed in the decay modes 
${\rm \bar{B}^0} \to {\rm D}^{*+}\ell^-{\rm X}$ and
${\rm B^-} \to {\rm D}^0\ell^-{\rm X}$ by
selecting muon and electron candidates in association with 
a fully reconstructed ${\rm D^0}$ or ${\rm D}^{*+}$ meson.
The ${\rm D^0}$ and ${\rm D}^{*+}$ are reconstructed from charged tracks and
$\pi^0$'s in the decay channels listed in Table 1. 
The tracks are required to come from a common vertex, and additional quality 
cuts on the momenta of the ${\rm D}^0$ candidates and the invariant mass of the 
${\rm D^{(*)}}\ell$ system
are applied in order to improve the signal to background ratio and 
to ensure well measured decay lengths. For the ${\rm D^{*+}}$ candidates,
the difference in mass between the reconstructed ${\rm D^{*+}}$ and the 
subsequent ${\rm D}^0$ is required to be within twice the 
experimental resolution of the nominal value of $145.4\;{\rm MeV}/c^2$.
Events reconstructed within two standard deviations of the fitted
${\rm D}^0$ mass are then selected for the lifetime analysis,
resulting in 1880 ${\rm D}^{*+}\ell^-$ and 2856 ${\rm D}^0\ell^-$
combinations. 

The decay length is calculated for these events from
the distance between the primary and B decay vertices, projected onto
the direction defined by the momentum of the ${\rm D}^{(*)}\ell$ system.
To reconstruct the momentum of the B meson, an energy flow technique 
is used, taking into account the missing energy from the undetected neutrino.

\begin{table}[htbp]
\tcaption{Number of ${\rm D}^0$ candidates and fraction of 
background events in the signal mass window.}
\begin{center}
\footnotesize\smalllineskip
\begin{tabular}[t]{ll c c}
        \hline 
        &  & \hspace*{3.2cm} & \hspace*{3.2cm} \mbox{}\\[-2ex]
        \multicolumn{2}{c}{{Subsample}}   
                    & {Candidates}         & 
                      {Background fraction} \\[1ex]   
        \hline 
	& & & \mbox{}\\[-2ex]
        {		 $\rm D^{*+} \ell^-$}      
                                & $\rm D^0 \to K^-\pi^+$        
                                & 651   & $0.066 \, \pm \, 0.004$ \\
                                & $\rm D^0 \to K^-\pi^+\pi^-\pi^+$      
                                & 670   & $0.096 \, \pm \, 0.004$  \\
                                & $\rm D^0 \to K^-\pi^+\pi^0$   
                                & 394   & $0.127 \, \pm \, 0.008$ \\
                                & $\rm D^0 \to K^0_S\,\pi^+\pi^-$       
                                & 165   & $0.061 \, \pm \, 0.006$ \\[1ex]
        \hline
	& & & \mbox{}\\[-2ex]
        {		 $\rm D^{0} \ell^-$} 
                                & $\rm D^0 \to K^-\pi^+$        
                                & 1312  & $0.133 \, \pm \, 0.006$ \\
                                & $\rm D^0 \to K^-\pi^+\pi^-\pi^+$      
                                & 664   & $0.232 \, \pm \, 0.012$ \\
                                & $\rm D^0 \to K^-\pi^+\pi^0$   
                                & 563   & $0.258 \, \pm \, 0.012$ \\
                                & $\rm D^0 \to K^0_S\,\pi^+\pi^-$       
                                & 317   & $0.139 \, \pm 0.009$ \\[1ex] 
        \hline 
\end{tabular}
\end{center}
\end{table}

\noindent

\section{Lifetime Measurement}
\noindent
The ${\rm \bar{B}}^0$ and ${\rm B}^-$ lifetimes are extracted using an
unbinned likelihood fit. Because the ${\rm D^0}\ell^-$ and ${\rm D}^{*+}\ell^-$
events contain a mixture of ${\rm {\bar B}^0}$ and ${\rm B}^-$, a 
simultaneous fit is performed to the proper time distributions
of both event samples.
Important inputs for this measurement are the branching ratios
for decay modes involving higher excited charm states,
${\rm B} \to {\rm D}^{**}\ell\nu({\rm X})$, which spoil the
${\rm \bar{B}}^0$ and ${\rm B}^-$ purity of the respective
${\rm D^0}\ell^-$ or ${\rm D}^{*+}\ell^-$ samples.
For the evaluation of the cross contamination, the
most recent ALEPH and DELPHI results \cite{BrALEPH,BrDELPHI} for
these branching ratios are used, leading to a significant reduction in the 
resulting uncertainty compared to a previous analysis \cite{OldPaper}.
The fit yields $\tau_{\rm \bar{B}^0}=1.518\pm0.053\;{\rm ps}$
and $\tau_{\rm B^-}=1.648\pm0.049\;{\rm ps}$, where the errors are statistical
only, and the statistical correlation is $-0.35$. The ratio of the lifetimes is 
found to be $\tau_{\rm B^-}/\tau_{\rm \bar{B}^0}=1.085\pm0.059$,
taking into account the correlation. The proper time distributions of
the two samples are shown in Fig.~1, with the results of the fit
superimposed.
Systematic uncertainties due to detector and physics modeling,
as well as those related to the fitting procedure,
are summarized in Table~2.

\begin{table}[htbp]
\tcaption{Summary of systematic uncertainties evaluated on the fitted lifetimes.}
\begin{center}
\footnotesize\smalllineskip
\begin{tabular}[t]{ l c c c }
        \hline 
	& & \mbox{}\\[-2ex]
        Source
        & $\quad\quad \tau_{\rm \bar{B}^0} (\rm ps) \quad\quad$ & $ \quad\quad \tau_{\rm B^-} (\rm ps)\quad\quad$ &
        $\quad \tau_{\rm B^-} / \tau_{\rm \bar{B}^0} \quad $ \\[1ex]
        \hline
	& & \mbox{}\\[-2ex]
        B momentum reconstruction       & $\pm 0.025$    &
$\pm 0.026$      & $\pm 0.009$      \\ 
        Background treatment            & $\pm 0.020$      & $\pm
0.020$  & $\pm 0.010$      \\
        Sample compositions             & $\pm 0.003$      & $\pm
0.003$  & $\pm 0.004$      \\
        $\rm D^{(*)}\pi\ell^-\nu$ relative efficiency
                                        & $\pm 0.006$      & $\pm
0.006$  & $\pm 0.006$      \\
        Decay length resolution         & $\pm 0.008$      & $\pm
0.008$  & $\pm 0.008$      \\[1ex]
        \hline
	& & \mbox{}\\[-2ex]
        Total                           & $\pm 0.034$    &
$\pm 0.035$      & $\pm 0.018$      \\[1ex]
        \hline
\end{tabular}
\end{center}
\end{table}

\begin{figure}[htb]
\centerline{\epsfig{figure=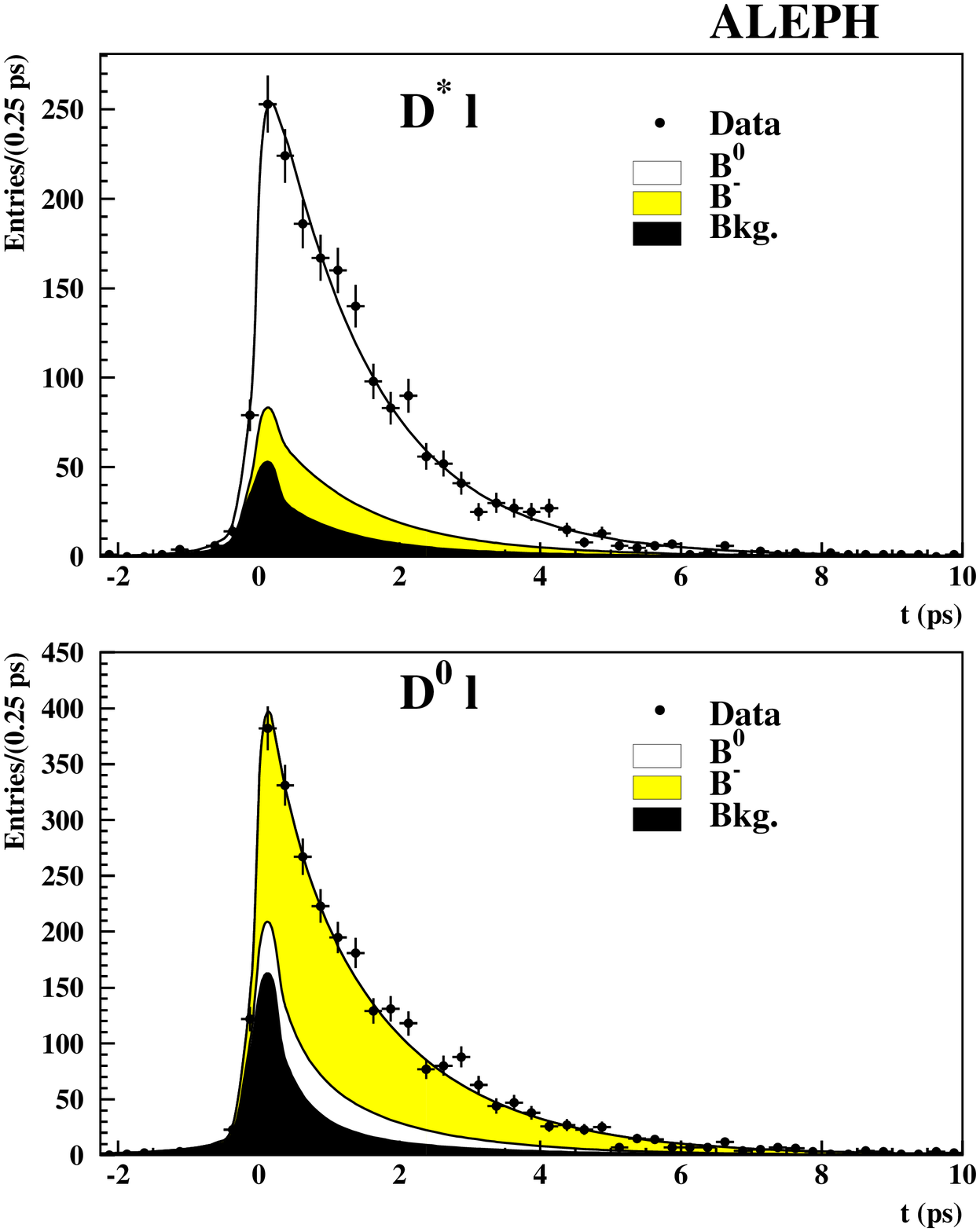, width=0.6\textwidth}}
\vspace*{3pt}
\fcaption{Proper time distributions for the ${\rm D}^{*+}\ell^-$ and
${\rm D}^0\ell^-$ samples, with the results of the fit superimposed.
Also shown are the background contributions and the respective 
${\rm {\bar B}^0}$ and ${\rm B}^-$ components.}
\end{figure}

\pagebreak
\section{Summary}
\noindent
The lifetimes of the charged and neutral B mesons have been measured using
data gathered with the ALEPH detector at LEP from 1991 to 1995.
A likelihood fit to the proper time distributions of 1880 
${\rm D}^{*+}\ell^-$ and 2856 ${\rm D^0}\ell^-$ candidates yields
the following results for the ${\rm \bar{B}}^0$ and ${\rm B}^-$ lifetimes
and their ratio:
$\tau_{\rm \bar{B}^0}=1.518\pm0.053\pm0.034\;{\rm ps}$, 
$\tau_{\rm B^-}=1.648\pm0.049\pm0.035\;{\rm ps}$, 
$\tau_{\rm B^-}/\tau_{\rm \bar{B}^0}=1.085\pm0.059\pm0.018$.

\nonumsection{Acknowledgments}
\noindent
I am grateful to my colleagues in ALEPH, in particular Giovanni~Calderini 
and Fabrizio~Palla, for providing the results presented in this talk.
It is a pleasure to thank the organizers for an enjoyable conference.

\nonumsection{References}

\end{document}